\documentclass[aps,pra,twocolumn,epsfig,amsfonts,amsmath,amstex,floatfix]{revtex4}

\usepackage{bbold}
\usepackage{color}

\usepackage{amsfonts,amsmath}
\usepackage{epsfig,amsmath}
\usepackage{graphicx}
\usepackage{dcolumn}
\usepackage{bm}
\usepackage{bbold}
\newcommand{\beq}{\begin{equation}}
\newcommand{\eeq}{\end{equation}}
\newcommand{\beqa}{\begin{eqnarray}}
\newcommand{\eeqa}{\end{eqnarray}}
\newcommand{\lam}{\lambda}

\newcommand{\la}{\langle} 
\newcommand{\ra}{\rangle}

\newcommand{\rp}{r_\perp} 

\def\cp#1{{ Contemp. \ Phys.} {\bf #1}}

\def\njp#1{{ New\ J.\ Phys.} {\bf#1}}

\def\oe#1{{ Opt.\ Express} {\bf#1}}

\def\ol#1{{ Opt.\ Lett.} {\bf#1}}
\def\opt#1{{ Optica} {\bf#1}}

\def\pra#1{{ Phys.\ Rev. A\/} {\bf#1}}

\def\prl#1{{ Phys.\ Rev.\ Lett.} {\bf#1}}

\def\pscr#1{{ Phys.\ Scrip.} {\bf#1}}

\def\rmp#1{{ Rev. \ Mod. \ Phys.} {\bf#1}}

\begin{document}

\title{Quantification and Observation of Genuine Three-Party Coherence – Classical Optics Has a Solution}
\author{X.-F. Qian$^{1}$}
\email{xqian6@stevens.edu}
\author{S. A. Wadood$^{2,3}$}
\author{A. N. Vamivakas$^{2,3,4}$}
\author{J. H. Eberly$^{2,3,4}$}
\affiliation{$^{1}$Department of Physics, and Center for Quantum Science and Engineering, Stevens Institute of Technology, Hoboken, New Jersey 07030, USA\\
$^{2}$Center for Coherence and Quantum Optics,
University of Rochester,
Rochester, New York 14627, USA\\
$^{3}$The Institute of Optics, University of Rochester, Rochester, New York 14627, USA\\
$^{4}$Department of Physics \& Astronomy, University of Rochester,
Rochester, New York 14627, USA}

\date{\today }

\begin{abstract}
We introduce a quantification of genuine three-party pure-state coherence for wave fields, classical and quantum, by borrowing concepts from classical optics. The tensor structure of a classical paraxial light beam composed of three principle degrees of freedom is shown to be equivalent to that of a three-qubit quantum state. The traditional basis-independent optical coherence quantity called degree of polarization is then determined to be the desired quantitative two-party coherence measure. When appropriately generalized, a set of fundamental constraint relations is derived among three two-party coherences. The constraint relations can be geometrically interpreted and visualized as tetrahedra nested within a coherence cube. A novel measure of three-party coherence is defined based on the constraints. We are reporting completed experimental tests and confirmations of the constraints as well as measurement of three-party coherence in the optical context. Our approach based on classical optics also opens a new way to analyze quantum coherence.
\end{abstract}


\maketitle

\noindent{\bf Introduction:} \quad It is sometimes remarked that one of the features showing the departure of quantum mechanics from the classical world is the coherent superposition of states \cite{Leggett}, but this overlooks the essential role that state superposition plays in every linear classical field theory, with electromagnetism as a chief example. A classical linear field theory is restricted by the same principles of linear vector space algebra as is quantum theory. Moreover, coherence itself is a concept that arose in connection with classical wave theory, principally in optics from the times of Huygens, Young and Fresnel (e.g., see \cite{Young, B&W}). The well known historical development and understanding of classical optics has produced one familiar family of coherence quantifications \cite{B&W}, focused on correlations within a single degree of freedom (e.g., temporal coherence or  spatial coherence - see \cite{statopt}). Recent studies of classical optical entanglement \cite{Spreeuw1998a, Ghose2001, Lee-Thomas2002, Simon-etal, Borges2010a, QE-OL, Kagalwala-etal-13, Toppel2014a, Zela2014a, Svozilik2015a, Khoury-OL, Ndagano2017a} have called attention to the desirability of exploring optical coherence across multiple degrees of freedom \cite{EberlyCP}. On the contrary, in the quantum realm the issue of coherence quantification was largely overlooked at first. Proposals for quantification of quantum mechanical coherence have very recently been made (see \cite{Plenio,  Mintert, Aberg, Girolami, Adesso, RMP}).  Coincidentally, in both quantum and classical contexts,  attention to details associated with more than two-party coherence is still developing, and experimental records of multi-party coherence quantification have been lacking.

Accompanying the recent search for coherence measures, two key issues remain largely open: (a) characterization(s) of genuine multiparty coherence and (b) specification(s) of constraints on multiple coherences. In the general case quantum states are multi-vector-space tensors, a feature widely shared among classical wave fields (see examples in optics: \cite{Gori, EberlyPhysScr}), which emphasizes the need for analyses of multiple and multiparty coherence. In the search for genuine multiparty measures, connections among subgroup quantities and their contribution to the overall multiparty structure are key elements \cite{Horodecki2007, Toth2009, Ali-Rau}.  A prominent example is the definition by Coffman, Kundu and Wootters \cite{CKW} of genuine three-qubit entanglement via a fundamental constraint called monogamy among participating subgroup two-party entanglements. Preliminary interesting characterizations of multiparty coherence have been proposed with entropic measures that apply directly to a multiparty quantum state by Yao et al., \cite{Yao} and Radhakrishnan et al., \cite{Byrnes}. Such definitions are independent of the structure of internal subgroup coherences, thus representing  approaches that differ from genuine multiparty structural measures.

As a first step toward genuine multiparty coherence, we focus here on the issue of quantifying and observing three-party pure-state coherence. This requires first a quantification of two-party coherences and then the identification of intrinsic restrictions among them. Despite that a number of quantifications of two-party coherence have been proposed \cite{RMP} for quantum states, measures that are feasible for the analysis of intrinsic three-party constraint are limited. Interestingly, recent studies of two-party coherences \cite{QE-OL, QMVE} in classical optical fields have provided an important clue to resolve the issue.

In the following we begin with a reminder of the tensor structural equivalence of a classical paraxial optical beam with that of a three-qubit quantum state. The conventional optical degree of polarization coherence (recognized as a two-party quantity \cite{QE-OL}) is employed to explore fundamental subgroup measures and restrictions. We do not enter the topic of higher-order quantum polarization properties, comprehensively treated by S\"oderholm, et al. \cite{Soderholm}. We can identify a novel set of constraint relations among three two-party coherences with a representation of visualizable geometric structures. A measure of genuine three-party coherence is then defined naturally, based on the constraints. Confirmation of these coherence constraints and observation of the proposed genuine three-party coherence are finally both presented experimentally in the classical optical context. 

We will adopt an attitude toward coherence itself, in the following sense. Recent work on the quantum side has been approached with an ``operational attitude", which is natural in regard to potential applications of coherence as an information resource \cite{RMP}. Consequently, an operation dependence that leads to basis dependence has been recognized as applying to the quantifications adopted with that attitude. Here we begin with an ``intrinsic content" attitude, the view that coherence refers to an intrinsic property of a field or state. This view implies that coherence refers to something about properties that are inherent, e.g., the ability to exhibit interference \cite{Young2}, independent of a basis chosen by an external party to undertake observation. Measurement of interference, in a particular basis, is only a way to exploit some or all of the available coherence. The fact that a coherence measurement is basis-dependent suggests a limitation of the observer or the operational scheme, not a limitation on intrinsic coherence. We note that a long-known optical measure, degree of polarization \cite{B&W}, is already a suitable basis-independent measure of vector coherence for consideration under this point of view. \\

\noindent{\bf Tensor structure equivalence:} Before addressing coherence quantification, we first remind that the tensor structure of a paraxial optical beam is equivalent to that of an arbitrary three-qubit quantum pure state. This will justify the universality of our results for both quantum and classical wave fields. 

A general three-qubit ($a, b, c$) pure state is given by
\beqa \label{3qubit}
|\psi\ra= \sum_{i,j,k=0,1}d_{ijk}|i_a\ra\otimes|j_b\ra\otimes|k_c\ra,
\eeqa
where $|i_a\ra$ represents qubit states $|0\ra$ and $|1\ra$ for qubit $a$, etc., and $d_{ijk}$ are the normalization coefficients.

The optical field of a general classical paraxial optical beam is written
\beqa \label{EBeamfcn}
 \vec E(\rp, t)  &=&   \hat x E_x(\rp,t) + \hat y E_y(\rp, t) \nonumber \\
& = & E_0[\alpha \hat{x}G_x(\rp)F_x(t)+\beta \hat{y}G_y(\rp)F_y(t)],
\eeqa
where its three independent physical properties (degrees of freedom) are: (i) the transverse polarization vectors $\hat{x}, \hat{y}$, (ii) the normalized transverse space functions $G_x(\rp), G_y(\rp)$ with $\int G^*_{\mu}(\rp)G_{\mu}(\rp)d\rp =1$ and $\mu=x,y$, and (iii) the normalized temporal functions $F_x(t),F_y(t)$ with $\la F_{\mu}^*( t)F_{\mu}(t) \ra=1$. The factorization indicated of the time and space dependences, i.e., $E_x(\rp,t)=G_x(\rp)F_x(t)$, etc., is available given the nature of the paraxial approximation \cite{paraxial}. Here $E_0$ is the overall amplitude of the beam with $ E_0^2=\la \vec E^*(\rp, t)\cdot \vec E(\rp, t) \ra =I$ representing the total intensity. The coefficients $\alpha$, $\beta$ are the normalized amplitudes corresponding to the transverse polarization components $\hat{x}$, $\hat{y}$ respectively with $|\alpha|^2+|\beta|^2=1$.

Along with the familiar vector nature of the orthogonal polarizations $\hat{x}, \hat{y}$, both the spatial and temporal functions are members of infinite dimensional vector spaces, spanned by complete sets of spatial and temporal modes \cite{B&W}. Thus, $G_{\mu}(\rp)$ and $F_{\mu}(t) $ are unit vectors in their corresponding vector spaces. Following the approach in Refs.~\cite{QE-OL, EberlyPhysScr}, we adopt Dirac notation and rewrite the optical beam (\ref{EBeamfcn}) as
\beq \label{EBeam}
|E\ra  = E_0 \big[\alpha  |x\ra \otimes|G_x\ra \otimes|F_x\ra+ \beta |y\ra\otimes |G_y\ra \otimes |F_y\ra  \big],
\eeq
and we will mostly omit tensor product symbols hereafter. The Dirac notation is only an indication of the mathematical vector nature, independent of being quantum or classical.

In the most general case, the $x$ and $y$ pairs of spatial and temporal functions are not mutually orthogonal, thus allowing arbitrarily assignable overlaps: $\la G_x|G_y\ra=\int G^*_{x}(\rp)G_{y}(\rp)d\rp = \delta$ and $\la F_x|F_y\ra =\la F^*_x(t)F_y(t)\ra=\gamma $ where both $|\delta | \le 1$ and $|\gamma | \le 1$ are guaranteed by the Schwarz inequality. This allows the two spatial vectors $|G_x\ra$, $|G_y\ra$ to be described by arbitrary superpositions of the two orthogonal bases, e.g., $|G_x\ra=p_x|G_0\ra+q_x|G_1\ra$, $|G_y\ra=p_y|G_0\ra+q_y|G_1\ra$, where $|p_{\mu}|^2+|q_{\mu}|^2=1$, $\la G_0|G_1\ra=0$ and the coefficients $p_x, q_x$ are independent of $p_y, q_y$. The same is true for the two temporal components which can be re-expressed as $|F_x\ra=r_x|F_0\ra+s_x|F_1\ra$, $|F_y\ra=r_y|F_0\ra+s_y|F_1\ra$. 

We will now denote each of the three vector spaces of the optical beam with a letter label for convenience: the traditional polarization vector components $|x\ra, |y\ra$ form a two-dimensional space we label as $a$, the spatial function vectors $|G_x\ra, |G_y\ra$ occupy another one labelled $b$, and the temporal function vectors $|F_x\ra, |F_y\ra$ occupy the third space labelled $c$. 

The replacement of $|G_x\ra$, $|G_y\ra$, $F_x\ra$, $|F_y\ra$ with the basis states $|G_0\ra$, $|G_1\ra$, $|F_0\ra$, $|F_1\ra$, allows us to see that the tensor structure of an arbitrary paraxial light beam (\ref{EBeam}) is equivalent to that of the arbitrary three-qubit quantum pure state (\ref{3qubit}). This fact allows the discussion of coherence quantification to be interchangeable between the two systems. \\



\noindent{\bf Multiple two-party polarization coherences:} Now we consider two-party coherence measures for the general paraxial optical beam (\ref{EBeam}), or equivalently the arbitrary three-qubit state (\ref{3qubit}). Recent studies of optical entanglement and its connection with the traditional polarization coherence show that both quantities are indeed two-party features \cite{QE-OL, QMVE}. That is, they describe a correlation (connection) between two or two groups of physical properties in a light field. Consequently, the so-called ``hidden coherences" were exposed in a multiparty context \cite{QMVE}.

The term ``polarization" is appropriate in more than one context in physics, and it always indicates a relationship between two independent attributes of a physical system (as highlighted in \cite{PCT}). Historically in optics ``degree of polarization" has measured the concentration or the ``alignment" of one attribute against the other, and the term ``completely polarized" means that the optical field exhibits a perfect alignment. This is in the sense that the entire optical amplitude aligns with a single vector direction, say $| u\ra$, in experimental lab space. It is the same as the expression 
\beq \label{Defalign}
|E'\ra= | u\ra | E_u(\rp, t) \ra
\eeq 
for the transverse optical field vector. That is, one needs the discrete two-dimensional vector space $\{|x\ra, |y\ra\}$ for $| u\ra$ as one party, and continuous linear $\{ {l}_2\}$ function spaces for $\rp$ and $t$ together as another party, for the total field amplitude  $|E_u(\rp, t)\ra$, and they are thus two-party factorable, tensor-separable, as shown in the field state $|E'\ra$ in (\ref{Defalign}). 

We will equate a fully polarization-coherent field with complete tensor separability in this way. Moreover, going forward we will employ the degree of polarization coherence, which can be interpreted as a degree of two-party separability ${\cal S}$. In the following we will see that paraxial beams, with their several vector-space degrees of freedom, enjoy several distinct coherence quantifications in terms of separability ${\cal S}$, including the quantification suggested by the traditional degree of polarization. We also remark that such coherence measures apply to the quantum state (\ref{3qubit}) directly due to the tensor structure equivalence.


To achieve the conventional degree of polarization coherence of the general optical beam (\ref{EBeam}), we will employ the outer product projector $|E\ra\la E|$ to obtain ${\cal W}$, the classical field's equivalent of a density matrix. We divide by $|E_0|^2$ to obtain the unit normalized form: 
\beq \label{W Def}
 {\cal W} = \frac{|E\ra\la E|}{|E_0|^2}.
 \eeq
A first analytical step is to trace over the $b$ and $c$ spaces, leading to a $2\times 2$ unit-normalized reduced density matrix in the $a$ space:
\beqa \label{a-cohmatrix}
{\cal W}_a &=&  \left[\begin{matrix}
{\cal W}_{xx} & {\cal W}_{xy} \\
{\cal W}_{yx} & {\cal W}_{yy}
\end{matrix} \right] 
=  \left[\begin{matrix}
|\alpha|^2 &\alpha\beta^* \delta^*\gamma^* \\
\alpha^*\beta \delta\gamma  &|\beta|^2 
\end{matrix} \right].
\eeqa
This is traditionally \cite{B&W} called the optical polarization coherence matrix,  which allows the long-known expression for the field's ``degree of polarization"  \cite{Wolf2007} to be interpreted as space $a$'s separability:
\beq \label{P_a def}
{\cal S}_a = \sqrt{1 - \frac{4Det{\cal W}_a}{(Tr{\cal W}_a)^2}}.
\eeq
This coherence quantity ${\cal S}_a$ is independent of any particular choice of basis, e.g., any basis in the $\{|x\ra,|y\ra\}$ space that's chosen to write $|E\ra$. This is assured because of the basis independence of determinant and trace.

An equivalent approach defines  ${\cal S}_a$ in terms of the eigenvalues $\{\lam_1, \lam_2\}$ of the ${\cal W}_a$ matrix:
\beq \label{P_a altdef}
{\cal S}_a = \Big|\frac{\lam_1 - \lam_2}{\lam_1 + \lam_2} \Big|.
\eeq
In either case the degree of polarization coherence for the $a$ space is obtained as
\beq \label{aSep}
{\cal S}_a = \sqrt{1-4|\alpha|^2|\beta|^2 [1 -| \delta\gamma|^2 ]}.
\eeq
The nature of the definition ensures that $1 \ge {\cal S} \ge 0$. 

As noted, separability is a two-vector-space property and such an optical beam (\ref{EBeam}) occupying three vector spaces can always be treated as a two-space structure by merging any two of the three spaces into one larger space. In this case of ${\cal S}_a$, the $b$ and $c$ spaces were treated together as a single space distinct from space $a$. 

One can obtain all three generic separability coherences (see related observations in \cite{QMVE}) in the three-space structure of $|E\ra$ and ${\cal W}$. However, to do this systematically for vector space $b$ we must first specify two orthogonal unit vectors in the $b$ space similar to $\{|x\ra, |y\ra \}$ in $a$ space. Instead of using the generic vectors $\{|G_0\ra, |G_1\ra\}$, we can accommodate this by introducing a new unit vector $|\bar G_x\ra$, explicitly defined to be orthogonal to $|G_x\ra$, allowing $|G_y\ra$ to remain unit-normalized when written
\beq
|G_y\ra = \delta |G_x\ra + \sqrt{1-|\delta|^2}\ |\bar G_x\ra,
\eeq
so that $\la G_x|G_y\ra = \delta$, as already specified below ({\ref{EBeam}), and  $|\bar{G}_x \ra$ is unit-normalized  and orthogonal to $|G_x\ra$, i.e.,$\la G_x|\bar G_x\ra$ = 0. Then the paraxial beam (\ref{EBeam}) can be rearranged as
\beqa 
\frac{|E\ra}{E_0}  =  \big[\alpha |x\ra |F_x\ra &+& \beta \delta |y\ra |F_y \ra \big] |G_x\ra \nonumber\\
&+& \big [\beta  \sqrt{1-|\delta|^2} |y\ra |F_y\ra \big] | \bar{G}_x\ra.
\eeqa
The entire continuous-variable transverse coordinate's vector space has been mapped onto a two-dimensional space spanned by the orthogonal basis $\{G_x(\rp), \bar{G}_x(\rp)\}$. Now one can conveniently trace over the $a$ and $c$ spaces to obtain ${\cal W}_b$, the coherence matrix for coherence space $b$ analogous to ${\cal W}_a$ in (\ref{a-cohmatrix}) in the form:
\beqa \label{b-cohmatrix}
{\cal W}_b &=& \left[\begin{matrix}
|\alpha|^2+|\beta\delta|^2 &|\beta|^2 \delta \sqrt{1-|\delta|^2} \\
|\beta|^2 \delta^* \sqrt{1-|\delta|^2} &|\beta|^2 (1-|\delta|^2)
\end{matrix} \right].
\eeqa
The separability $S_b$ subsequently provides the following quantification of coherence between $b$ space and the remaining two:
\beqa
{\cal S}_b &=& \sqrt{1-4 |\alpha\beta|^2 (1-|\delta|^2)}.
\eeqa

Similarly, the component states $|F_x\ra$ and $|F_y\ra$ for the temporal function $F(t)$ are in general not orthogonal, but an orthogonal basis state can be found in the same way, so $|F_y\ra$ can be re-expressed as 
\beq
|F_y\ra =\gamma |F_{x}\ra + \sqrt{1-|\gamma|^2} |\bar{F}_x \ra,
\eeq
so that both $\la F_y|F_y\ra$ = 1 and $\la F_x|F_y\ra = \gamma$ are satisfied. Consequently, the corresponding coherence matrix and degree of coherence in $c$ space are obtained as 
\beqa \label{c-cohmatrix}
{\cal W}_c &=& \left[\begin{matrix}
|\alpha|^2 + |\beta\gamma|^2 &|\beta|^2 \gamma \sqrt{1- |\gamma|^2} \\
|\beta|^2 \gamma^* \sqrt{1- |\gamma|^2} &|\beta|^2 (1- |\gamma|^2)
\end{matrix} \right],
\eeqa
and 
\beqa
{\cal S}_c  & = &  \sqrt{1 - 4 |\alpha\beta|^2 (1 - |\gamma|^2)}.
\eeqa

This completes the derivation of the three coherences ${\cal S}_a$, ${\cal S}_b$, ${\cal S}_c$ for a typical paraxial beam comprising three degrees of freedom and occupying three vector spaces. The three coherences apply directly to the quantum three-qubit state (\ref{3qubit}), corresponding to each qubit respectively. It is important that these coherences are distinct but not independent, as we show next.\\

\noindent{\bf Coherence restrictions:} Each of the three coherence measures is bounded in the same way: $0 \le {\cal S}_a, {\cal S}_b, {\cal S}_c \le1$. The sum of any two coherences, e.g., $ {\cal S}_a+ {\cal S}_b$, is thus bounded by 0 and 2, and the sum of all three lies between 0 and 3. One might expect that the difference between any two-coherence sum and the remaining coherence should be bounded by 2, but the generic structure of the light beam (\ref{EBeam}), equivalently the three-qubit state (\ref{3qubit}), forces this restriction to become much tighter: 
\beq
|{\cal S}_{a} \pm {\cal S}_{b} - {\cal S}_{c} | \le 1, \label{3inequality}
\eeq
quantifying the restrictions existing among the three coherences. Here $a,b,c$ can switch order freely. This set of symmetric inequalities is our first main result. It simply says that the sum of any two separabilities, upon subtracting the third, is always less than unity. A related set of inequalities for multiparty entanglement can be found in Refs.~\cite{AQE2016, QAE2018}. The detailed proof of this set of coherence inequalities is given in the Appendix.

\begin{figure}[h!]
\includegraphics[width=7cm]{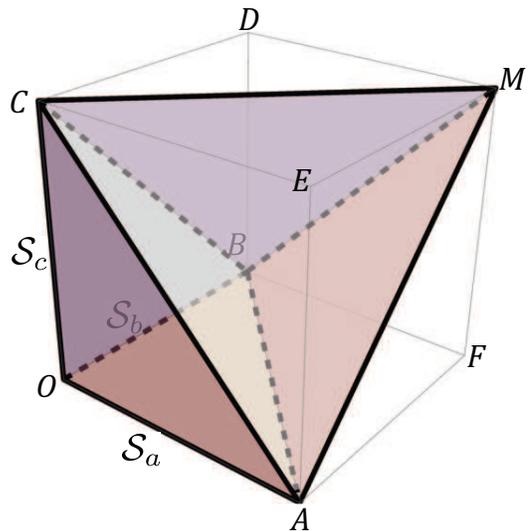}
\caption{Coherence tetrahedra. The allowed region restricted by (\ref{3inequality}) is represented by the two tetrahedra $OABC$ and $MABC$. Point $O$ corresponds to the origin with no coherence at all, i.e., $({\cal S}_a,{\cal S}_b,{\cal S}_c) = (0, 0, 0)$, and point $M$ represents maximum coherence with $({\cal S}_a,{\cal S}_b,{\cal S}_c)=(1, 1, 1)$.}
\label{cube}
\end{figure}

There is a geometric representation of algebraic relation (\ref{3inequality}) that provides a direct illustration of it. We define a natural separability coherence vector $\vec {\cal S} = ({\cal S}_a, {\cal S}_b, {\cal S}_c )$, and let the separabilities be the three axes of a unit cube, as in Fig.~\ref{cube}. When there is no restriction among the three coherences, the separability vector $\vec {\cal S}$ will occupy the entire cube. If a restriction applies, it will reduce the physically habitable region. For example, the inequality ${\cal S}_{b}+{\cal S}_{c} - {\cal S}_{a} \le 1$ excludes tetrahedron $BCMD$. Similarly, the remaining two relations in (\ref{3inequality}) exclude two corresponding tetrahedra $CAME$ and $ABMF$. The result is that  the allowed region for occupation is two tetrahedra $OABC$ and $MABC$ with the triangle $ABC$ as their common base (see Fig.~\ref{cube}). 


Notice that all the points on a given plane that is perpendicular to the body diagonal line $OM$ (e.g., $\triangle ABC$) have the same total separability: ${\cal S}_{\rm total} = {\cal S}_{a} + {\cal S}_{b} + {\cal S}_{c}$. This provides the opportunity to consider the flexibility that a given total amount has in being distributed. Its quantification can be geometrically represented by the area of triangles that are perpendicular to the body diagonal line $OM$ (and which live inside the confined domain defined by the two tetrahedra $OABC$ and $MABC$). See a similar analysis in \cite{QAE2018} for entanglement.\\

\noindent{\bf Genuine three-party coherence:} The individual separabilities ${\cal S}_a,{\cal S}_b,{\cal S}_c$ are two-party (two degrees of freedom) coherences \cite{QE-OL}. The distribution and sharing restriction analysis described in the previous section provides an optimal platform for the consideration of genuine three-party coherence. From the restriction inequalities, three ``directed" coherence quantities can be defined, i.e.,  
\beqa
\mathbb{C}_{a\rightarrow bc} &\equiv &1+{\cal S}_a - {\cal S}_b - {\cal S}_c\ge 0,\\
\mathbb{C}_{b\rightarrow ca} &\equiv &1+{\cal S}_b - {\cal S}_c - {\cal S}_a \ge 0,\\
 \mathbb{C}_{c\rightarrow ab} &\equiv &1+{\cal S}_c - {\cal S}_a - {\cal S}_b\ge 0.
\eeqa

Each quantity represents a directional residual coherence when one individual coherence is reduced by the sum of the other two. It is obvious that each directional coherence quantity involves all three paraxial degrees of freedom, thus representing a biased three-party coherence. As remarked at the beginning, we follow the traditional Coffman-Kundu-Wootters procedure \cite{CKW} for defining a genuine multiparty quantity \cite{genuine}. We find that an attractive measure of genuine three-party coherence of the beam is then obtained when all three directional quantities are non-zero. That is, the degree of genuine three-party coherence can be simply defined as the minimum of the three directional three-party coherences, i.e.,
\beq \label{3coherence}
\mathbb{C}_{abc}\equiv {\rm Min}\{\mathbb{C}_{a\rightarrow bc}, \mathbb{C}_{b\rightarrow ca}, \mathbb{C}_{c\rightarrow ab} \}.
\eeq

Now we show that this genuine three-party coherence measure is automatically normalized, i.e., $0\le \mathbb{C}_{abc}\le 1$. Suppose ${\cal S}_a \ge {\cal S}_b \ge {\cal S}_c$ (without loss of generality), then the minimal directional coherence is $\mathbb{C}_{c\rightarrow ab}$. This leads to the result
\beq
0\le \mathbb{C}_{abc}= 1+ {\cal S}_c - {\cal S}_a - {\cal S}_b \le 1- {\cal S}_a \le 1,
\eeq
where we have used the fact that ${\cal S}_c \le {\cal S}_b$. 

To have a better understanding of the meaning of the genuine three-party coherence $\mathbb{C}_{abc}$, we discuss three extreme cases in the following.

\begin{figure*}[t]
\includegraphics[width=13cm]{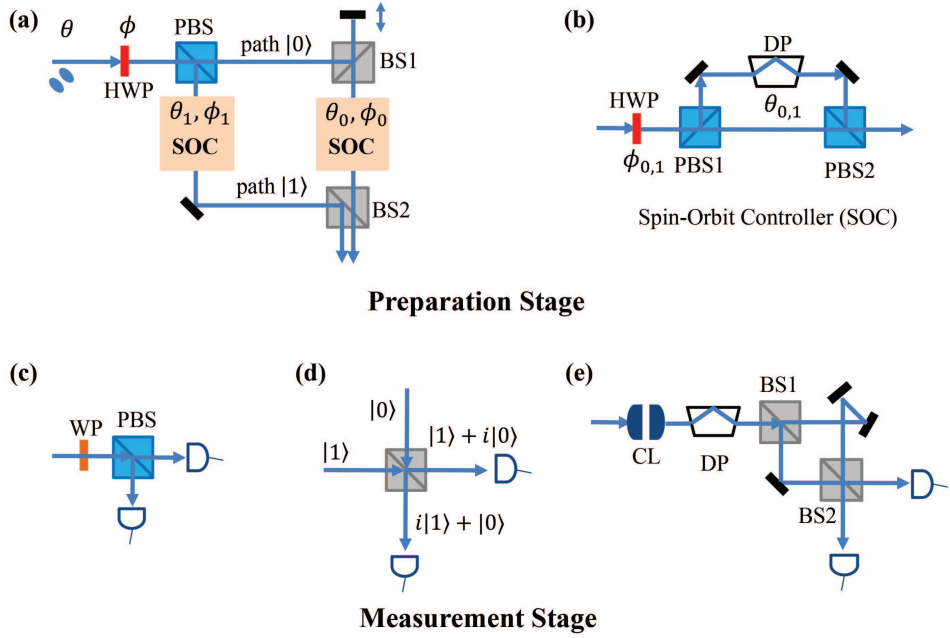}
\caption{Experimental setup. Panel (a) illustrates the modified Mach-Zehnder interferometer used to created general optical beams in the arbitrary structure (\ref{expbeam}). Panel (b) shows the detailed components of the spin-orbit controller. In the detection stage, panels (c), (d), (e) illustrate the tomographic measurement setups of coherence matrices ${\cal W}_a$, ${\cal W}_b$, ${\cal W}_c$ respectively. }
\label{setup}
\end{figure*}

(1) ${\cal S}_a = {\cal S}_b = {\cal S}_c =1$, which means all three degrees of freedom are completely coherent, i.e., each degree of freedom (DoF) is separable from the remaining two DoFs of the optical beam. Then all DoFs are independent of each other. Therefore, there should be no three party coherence, which is indicated by the measure $\mathbb{C}_{abc}=\mathbb{C}_{b\rightarrow ca}=\mathbb{C}_{c\rightarrow ab}=\mathbb{C}_{a\rightarrow bc}\equiv1+ {\cal S}_a -  {\cal S}_b - {\cal S}_c = 0$. 

(2) ${\cal S}_a = {\cal S}_b = {\cal S}_c = 0$, which means all three degrees of freedom are incoherent (non-separable). Although incoherent individually, this case brings maximal mutual dependence among the three DoFs. That is, three incoherent DoFs can behave coherently as a whole. Therefore, there should be maximum three-party coherence, and this is also directly shown by the measure $\mathbb{C}_{abc}$ taking its maximum value 1, i.e., $\mathbb{C}_{abc}=\mathbb{C}_{b\rightarrow ca}=\mathbb{C}_{c\rightarrow ab}=\mathbb{C}_{a\rightarrow bc}\equiv 1 + {\cal S}_a - {\cal S}_b - {\cal S}_c=1$. 

(3) ${\cal S}_a$ = ${\cal S}_b$, ${\cal S}_c$ = 1, which means DoF $c$ is fully coherent and the remaining two are equally partially coherent. When ${\cal S}_c = 1$, the $c$ DoF is completely independent of the remaining two, indicating zero-biased residual coherences. Therefore, there should be zero genuine three-party coherence, as indicated by the measure $\mathbb{C}_{abc}=\mathbb{C}_{b\rightarrow ca}=\mathbb{C}_{a\rightarrow bc}=1 + {\cal S}_a - {\cal S}_b - {\cal S}_c = 0$. 

These three extreme cases consistently confirm that the three-party coherence $\mathbb{C}_{abc}$ characterizes the coherence property of all three DoFs as an interdependent unit. Therefore it can be viewed as the coherence of the electromagnetic field as a whole \cite{Ari}. \\


\noindent{\bf Experimental confirmation:}  Now we describe an experimental test of the coherence restriction inequalities (\ref{3inequality}) and a measurement of the three-party coherence (\ref{3coherence}) with an optical beam. To accomplish these two tasks, it is first needed to produce an optical beam with the three-DoF structure described in (\ref{EBeam}) and then measure all three separability coherences with respect to each optical DoF. We remark that the procedure of testing can be similar for a quantum system (e.g., with single photons) in the three-qubit state (\ref{3qubit}), but of course with corresponding technical details.

The general field in (\ref{EBeam}) is provided by a laser beam with two polarization components, two spatial modes and two path modes. Specifically, the two polarization vectors in space $a$ are still described as $ |x\rangle $, $ |y\rangle $. The spatial components $|G_x\ra, |G_y\ra$ in space $b$ are represented by superpositions of two first-order Hermite-Gauss (HG) modes, $|G_{10}\rangle \equiv HG_{10}$, $ |G_{01}\ra \equiv HG_{01} $, i.e., $|G_x\rangle = c_1 |G_{10}\ra + c_2 |G_{01}\ra$ and $|G_y\rangle = c'_1 |G_{10}\ra + c'_2 |G_{01}\ra$, where the coefficients $c_1,c_2$ are independent of $c'_1,c'_2$.  For simplicity and without loss of generality, the temporal components $|F_x\ra $, $ |F_y\ra $ in space $c$ are replaced by the superpositions of two path modes $ |0\rangle $, $|1\rangle$, i.e., $|F_x\ra =d_0 |0\ra + d_1 |1\ra$ and $|F_y\ra =d'_0 |0\ra + d'_1 |1\ra$ with coefficients $d_0,d_1$ independent $d'_0,d'_1$. It is emphasized that such a replacement doesn't change the vector structure of the general three-DoF form described in Eq.~(\ref{EBeam}). 
 
A 795 nm laser is directed to a spatial light modulator (SLM) to create a beam with an arbitrarily oriented first-order HG mode $|G_\theta\ra=\cos\theta |G_{10}\ra+\sin\theta |G_{01}\ra$ with vertical polarization $|y\ra$. Then it enters the preparation stage, as shown in Fig.~\ref{setup} (a), to produce an arbitrary three-DoF structured beam. It first passes a half-wave plate (HWP) to change the polarization to an arbitrary orientation, i.e., $|\phi\ra=\cos\phi |x\ra+\sin\phi |y\ra$. Then the beam is split into two by a polarizing beamsplitter (PBS) where the transmission-reflection ratio depends on the rotation angle $\phi$ of the HWP in front of it. The two arms of the Mach-Zehnder interferometer represent the two paths $|0\ra, |1\ra$. A superposition of the two paths can be realized by combining the two with a 50/50 beamsplitter (BS2). When no superposition is needed, the two paths are combined with a transverse shift as shown in Fig.~\ref{setup} (a) - so that the two outputs are parallel without overlap.

The transmitted component travels in path $|0\ra$ to be directed by a 50/50 beamsplitter (BS1) to a movable mirror for an appropriate phase adjustment $\Delta$. Then it enters a spin-orbit controller (SOC) to combine with path $|1\ra$ at BS2. The reflected component from the PBS travels in path $|1\ra$ to enter another SOC before entering BS2.

An SOC is a modified Mach-Zehnder interferometer (see Fig.~\ref{setup} (b)) that manipulates the spatial modes of the beam depending on spin polarizations. For example, in path $|1\ra$, the incoming signal $|1\ra|y\ra|G_{\theta}\ra$ enters a HWP to become $|1\ra (\cos\phi_1 |x\ra+\sin\phi_1 |y\ra)  |G_\theta\ra$ with an arbitrary angle $\phi_1$. Then it enters the modified Mach-Zehnder interferometer, composed of two PBSs and a Dove Prism (DP). The spatial mode of the $|x\ra$ polarization remains unchanged while that of the $|y\ra$ component changes into an arbitrarily oriented mode controlled by the DP via a rotation angle $\theta_1$, i.e., $|G_{\theta_1}\ra=\cos\theta_1|G_{10}\ra+\sin\theta_1|G_{01}\ra$. So in path $|1\ra$, the output beam of the SOC can be described as
\beq
|e_1\ra=|1\ra \Big(\cos\phi_1|x\ra|G_{\theta}\ra+ \sin\phi_1 |y\ra|G_{\theta_1}\ra \Big).
\eeq
Similarly, the output beam of the SOC in path $|0\ra$ can be described as
\beq
|e_0\ra= |0\ra \Big(\cos\phi_0|x\ra|G_{\theta}\ra+ \sin\phi_0 |y\ra|G_{\theta_0}\ra \Big),
\eeq
where the corresponding HWP in path $|0\ra$ has produced the polarization $|\phi_0\ra=\cos\phi_0 |x\ra+\sin\phi_0 |y\ra$

As a result, the prepared beam at the output of BS2, see Fig.~\ref{setup} (a), is in the form 
\beq \label{expbeam}
\frac{|E\ra}{\sqrt{N}}=\sin\phi|e_1\ra+\frac{\cos\phi }{2}e^{i\Delta} |e_0\ra,
\eeq
where $N$ is the normalization factor, which is the intensity $I$. Since the rotation parameters $\theta$, $\theta_0$, $\theta_1$, $\phi$, $\phi_0$, $\phi_1$, and the relative phase factor $\Delta$ are independent of each other, the prepared beam (\ref{expbeam}) is a general representation of the three-DoF structure (\ref{EBeam}) and the three-qubit quantum state (\ref{3qubit}). 

To obtain the separability coherences ${\cal S}_a, {\cal S}_b, {\cal S}_c$, tomographic setups are employed to measure the corresponding coherence matrices (\ref{a-cohmatrix}), (\ref{b-cohmatrix}), (\ref{c-cohmatrix}) through Stokes-like parameters, analogous to qubit Pauli matrix parameters \cite{James-etal}. The Stokes basis projection for spin polarization is realized by the standard combination of (half- and quarter-) wave plates and a polarizing beamsplitter as shown in Fig.~\ref{setup} (c). The projection in the spatial-mode degrees of freedom is realized by the combination of a pair of cylindrical lenses, a Dove prism and a modified Mach-Zehender interferometer - see illustration in Fig.~\ref{setup} (e) and detailed description in Ref.~\cite{QMVE}. 

The projection in Stokes-like basis in the path degrees of freedom is realized by BS2 together with a translation stage in path $|0\ra$. In this case, the $|0\ra$ or $|1\ra$ basis is analyzed by simply blocking either one of the two paths. When the two paths are perfectly collimated at BS2, the two outputs are effectively in the basis $|0\ra\pm i |1\ra$, as shown in Fig.~\ref{setup} (d). The $|0\ra\pm  |1\ra$ basis is realized by a $\Delta=\pi/2$ phase delay in path $|0\ra$.

\begin{table*}[t] 
\begin{center}
\begin{tabular}{|r|r|r|r|r|r|r|r|r|r|} 
\hline
Beams & \ ${\cal S}_a \ \ $ &${\cal S}_b \ \ $&${\cal S}_c \ \ $& ${\cal S}_b+{\cal S}_b-{\cal S}_a$ & ${\cal S}_c+{\cal S}_a-{\cal S}_b $    & ${\cal S}_a+{\cal S}_b-{\cal S}_c$ & $\mathbb{C}_{abc}$ &  $\mathbb{C}^{T}_{abc}$& red dots     \\
\hline 
\hspace{0.5mm}
$|E_1\ra$  \ \ & 0.993 $\pm$ 0.015 & 0.988 $\pm $ 0.021& 0.995 $\pm$ 0.014 & 0.990 \ \ &  0.999 \ \  &  0.987 \ \ \  & 0.001 & 0 & M  \\
\hline 
\hspace{0.5mm}
$|E_2\ra$  \ \ & 0.015 $\pm$ 0.016  & 0.017 $\pm$ 0.028 & 0.013 $\pm$ 0.019 & 0.014 \ \    & 0.011 \ \ & 0.019 \ \ \  & 0.981 & 1 &  O  \\
\hline 
\hspace{0.5mm}
$|E_3\ra$  \ \ & 0.348 $\pm$ 0.022& 0.319$\pm$ 0.033 & 0.340$\pm$ 0.020 & 0.311\ \ \    & 0.368 \ \ &0.327 \ \ \  & 0.632 & 2/3& W   \\
\hline 
\hspace{0.5mm}
$|E_4\ra$  \ \ & 0.015 $\pm$ 0.009 & 0.024 $\pm$ 0.021 & 0.991 $\pm$ 0.012 & 1.000\ \ \    & 0.982  \ \ & $-0.952$ \ \ \  & 0.000 & 0&  A   \\
\hline 
\hspace{0.5mm}
$|E_5\ra$  \ \ & 0.017$\pm$ 0.024 & 0.983 $\pm$ 0.049& 0.014 $\pm$ 0.023& 0.980\ \ \    & $-0.953$   \ \ & 0.986 \ \ \  & 0.014 & 0&  B   \\
\hline 
\hspace{0.5mm}
$|E_6\ra$  \ \ & 0.989$\pm$ 0.032 & 0.026$\pm$ 0.052 & 0.019 $\pm$ 0.025 & $-0.944$\ \ \    & 0.983  \ \ &0.996 \ \ \  & 0.004& 0&  C   \\
\hline 
\end{tabular}
\end{center}
\caption{Measured values of coherences ${\cal S}_a$, ${\cal S}_b$, ${\cal S}_c$ and the corresponding genuine three-party coherence ${\cal C}_{abc}$ for six representative light fields. The theoretical three-party coherence ${\cal C}^{T}_{abc}$ is listed for comparison with the experiment. Three corresponding inequalities are calculated and displayed confirming the restriction relation (\ref{3inequality}). Connection of different beams to red dots of the tetrahedra in Fig.~\ref{expcube} are also established by the labeling letters.  } \label{Table}
\end{table*}

To cover all interesting properties of the coherence restriction inequalities (\ref{3inequality}), six specific representative light fields (\ref{expbeam}) are chosen for the measurements of all three coherences ${\cal S}_a, {\cal S}_b, {\cal S}_c$ and the three-party coherence $\mathbb{C}_{abc}$. These optical fields correspond to states with interesting entanglement properties as well. The first case we prepare is a completely separable tensor product of all three DoFs, i.e., 
\beq \label{E1}
|E_1\ra=|y\ra\otimes|G_{01}\ra\otimes |1\ra.
\eeq
This beam corresponds to complete polarization, i.e., full separability coherence, of each DoF with ${\cal S}_a={\cal S}_b={\cal S}_c=1$. To generate such a beam, the SLM is set to produce a $|G_{10}\ra$ ($\cos\theta=1$) mode with vertical polarization $|y\ra$. The HWP in Fig.~\ref{setup} (a) is set to leave the polarization state unchanged ($\cos\phi=0$). Then it is completely reflected by the PBS to path $|1\ra$ to enter the SOC, whose HWP is also set to keep the polarization state (i.e., $\cos\phi_1=0$). The reflected beam by PBS1 passes through a DP to change into vertically oriented first order Hermite-Gauss mode $G_{01}$ ($\cos\theta_1=0$). Then the output beam after BS2 is exactly characterized as $|E_1\ra$ in (\ref{E1}). 

The second beam we produce is a GHZ-type of entangled state with each individual DoF maximally dependent on the remaining two, i.e.,
\beq
|E_2\ra=|G_{01}\ra\otimes|y\ra\otimes |1\ra+|G_{10}\ra\otimes|x\ra\otimes |0\ra.
\eeq
It corresponds to the case when all DoFs are completely inseparable ${\cal S}_a={\cal S}_b={\cal S}_c=0$. It is realized with an incoming beam of $|G_{10}\ra$ ($\theta=0$) mode and vertical polarization $|y\ra$, by setting the rotation parameters of three HWPs to be $\cos\phi=\sqrt{4/5}$, $\cos\phi_1=0$, $\cos\phi_0=1$ respectively, and the rotation parameter the DP in path $|1\ra$ to be $\cos\theta_1=0$. 

The third beam to test is a W-type state, i.e.,
\beq
|E_3\ra=|G_{01}\ra\otimes|y\ra\otimes |0\ra+|G_{01}\ra\otimes|x\ra\otimes |1\ra+|G_{10}\ra\otimes|y\ra\otimes |1\ra.
\eeq
In this beam, all three DoFs are equally partially coherent. To generate $|E_3\ra$, the incoming beam is produced with $|G_{01}\ra$ ($\theta=\pi/2$) mode and vertical polarization $|y\ra$. The three HWPs are set to be $\cos\phi=\sqrt{4/6}$, $\cos\phi_1=\sqrt{1/2}$, $\cos\phi_0=0$, and the two DPs are set to be $\cos\theta_1=1$, $\cos\theta_0=0$.

\begin{figure}[t!]
\includegraphics[width=7cm]{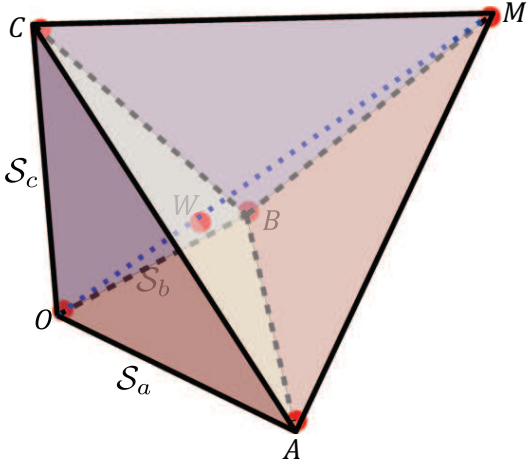}
\caption{Geometric representation of measured coherence values for all six prepared beams by the corresponding red dots. The three coordinates of each red dot represent the coherence values $({\cal S}_a, {\cal S}_b, {\cal S}_c)$ respectively.}
\label{expcube}
\end{figure}

We have also tested three more representative beams with two of the three DoFs in a Bell-type state and the third DoF completely separable with the first two, i.e.,
\beqa
|E_4\ra&=&|1\ra\otimes \Big(|G_{01}\ra\otimes |x\ra+|G_{10}\ra\otimes |y\ra \Big), \\
|E_5\ra&=&|G_{01}\ra\otimes \Big(|y\ra\otimes |0\ra+|x\ra\otimes |1\ra \Big), \\
|E_6\ra&=&|y\ra\otimes \Big(|G_{01}\ra\otimes |0\ra+|G_{10}\ra\otimes |1\ra \Big).
\eeqa
For these three beams, one of the three DoFs is completely coherent and the remaining two DoFs are completely incoherent. The three beams can be prepared with the same initial incoming beam with $|G_{01}\ra$ ($\theta=\pi/2$) mode and vertical polarization $|y\ra$. For state $|E_4\ra$, the HWPs are set as $\cos\phi=0$, $\cos\phi_1=\sqrt{1/2}$, and the DPs are set to be $\cos\theta_1=1$,  $\cos\theta_0=any$. For state $|E_5\ra$, the HWPs are set as $\cos\phi=\sqrt{4/5}$, $\cos\phi_1=1$, $\cos\phi_0=0$, and the DPs are set to be  $\cos\theta_1=any$, $\cos\theta_0=0$. For state $|E_6\ra$, the HWPs are set as $\cos\phi=\sqrt{4/5}$, $\cos\phi_1=0$, $\cos\phi_0=0$, and the DPs are set as $\cos\theta_1=1$, $\cos\theta_0=0$.

The measured three separability coherences ${\cal S}_a, {\cal S}_b, {\cal S}_c$ for the six prepared light beams are illustrated in Fig.~\ref{expcube} by the red dots. The three coordinates of each red dot represent the values ${\cal S}_a, {\cal S}_b, {\cal S}_c$ respectively for a corresponding beam. Here the light beams $|E_{1,2,3,4,5,6}\ra$ correspond to the red dots $M, O, W, A, B, C$ respectively. Specific values of each set of coherences $({\cal S}_a, {\cal S}_b, {\cal S}_c)$ are listed in detail in Table \ref{Table} for the six corresponding beams. The genuine three-party coherences $\mathbb{C}_{abc}$ are also obtained for the first time for each of the six beams. As is apparent, the measured results confirm the generic coherence restriction relation $|{\cal S}_a\pm {\cal S}_b-{\cal S}_c|\le 1$.\\

\noindent{\bf Summary and discussion:} \quad We have approached the issue of quantification of genuine three-party pure-state coherence from the view of intrinsic content. The equivalence of qubit-state tensor structure to that of a generic paraxial light beam has allowed us to resolve the overall issue with concepts and results from classical optics. By adopting the generalized basis-independence measure of optical polarization coherence, we have identified a set of symmetric coherence constraint relations that can be geometrically viewed as tetrahedral inequalities. These inequalities embody a fundamental multi-coherence law for paraxial light beams as well as for arbitrary three-qubit pure states. They lead to expression (\ref{3coherence}) for $\mathbb{C}_{abc}$, which we believe to be the first quantification of genuine three-party optical coherence (or three-qubit pure-state quantum coherence). It is different from the quantum entropic characterizations \cite{Yao,Byrnes} which are directly related to the multiparty state entropy.

We have further demonstrated the experimental confirmation of the tetrahedral coherence inequalities with three degrees of freedom of a paraxial laser beam: ordinary transverse polarization, spatial mode and independent path selection. The observation of genuine three-party coherence was consequently achieved and shown to be consistent with the constraints we derived. Our optical setup provides a useful platform to analyze comprehensively the coherence properties for all three degrees of freedom of light beams.

Our special classical optical approach along with geometric analysis on coherence opens a new systematic way to investigate and represent quantum coherence. The analysis applies straightforwardly to multiparty systems. It is expected that the coherence constraint inequalities and the associated multiparty coherence definition can provide useful new guidance for the sharing and distribution of coherence in various optical and quantum information tasks.\\


\noindent{\bf Appendix - Proof of Separability Coherence Restriction:} 
Here we prove the tetrahedral coherence restriction relation (\ref{3inequality}) via the eigenvalues of each of the normalized coherence matrices ${\cal W}_a, {\cal W}_b, {\cal W}_c$. By definition, the three corresponding separability coherences are given as 
\beqa
{\cal S}_a&=& \lambda_1^{(a)}-\lambda_2^{(a) },\\
{\cal S}_b&=& \lambda_1^{(b)}-\lambda_2^{(b)}, \\
{\cal S}_c&=& \lambda_1^{(c)} -\lambda_2^{(c) },
\eeqa
where we have assumed $ \lambda_1^{(\mu)}\ge \lambda_2^{(\mu) }$, $\mu=a,b,c$ without loss of generality, and have used the normalization condition  $ \lambda_1^{(\mu)}+ \lambda_2^{(\mu)}=1$.  

In the following we describe the proof of the inequality 
\beqa \label{inequality1}
1+ {\cal S}_a\ge{\cal S}_b+{\cal S}_c,
\eeqa
and the remaining two can be proved similarly.

The eigenvectors of each of the coherence matrices ${\cal W}_a$, ${\cal W}_b$, ${\cal W}_c$ can be defined correspondingly and denoted as $|\phi^{(a)}_1\ra$, $|\phi^{(a)}_2\ra$, $|\phi^{(b)}_1\ra$, $|\phi^{(b)}_2\ra$, $|\phi^{(c)}_1\ra$, $|\phi^{(c)}_2\ra$. Then one can rewrite the unit-normalized paraxial field as 

\beqa \label{decompose1}
\frac{|E\ra}{E_0} & = & \sqrt{ \lambda_1^{(a)}}|\phi^{(a)}_1\ra[x_1|\phi^{(b)}_1\ra|\phi^{(c)}_1\ra+x_2|\phi^{(b)}_1\ra|\phi^{(c)}_2\ra \notag \\
&&+ x_3|\phi^{(b)}_2\ra|\phi^{(c)}_1\ra+x_4|\phi^{(b)}_2\ra|\phi^{(c)}_2\ra ]  \notag  \\ 
&&+ \sqrt{ \lambda_2^{(a)}}|\phi^{(a)}_2\ra[y_1|\phi^{(b)}_1\ra|\phi^{(c)}_1\ra+y_2|\phi^{(b)}_1\ra|\phi^{(c)}_2\ra \notag \\
&&+ y_3|\phi^{(b)}_2\ra|\phi^{(c)}_1\ra+y_4|\phi^{(b)}_2\ra|\phi^{(c)}_2\ra ],
\eeqa
with the orthonormality conditions given as

\begin{subequations}
\begin{eqnarray}
\sum_{j=1}^{4} |x_{j}|^{2} & = & \sum_{j=1}^{4} |y_{j}|^{2}=1,
\label{condition1} \\
\sum_{j=1}^{4}x_{j} y_{j}^{\ast } &=& 0. 
\label{condition2}
\end{eqnarray}
\end{subequations}

When factoring out the states $|\phi^{(b)}_1\ra$ and $|\phi^{(b)}_2\ra$ of the field (\ref{decompose1}), it is straightforward to obtain eigenvalues for ${\cal W}_b$ in vector space $b$, i.e.,

\begin{eqnarray}
\lambda_1^{(b)} &=&\sum_{j=1,2}(\lambda
_{1}^{(a)}|x_{j}|^{2}+\lambda _{2}^{(a)}|y_{j}|^{2}), \\
\lambda_2^{(b)} &=&\sum_{j=3,4}(\lambda
_{1}^{(a)}|x_{j}|^{2}+\lambda _{2}^{(a)}|y_{j}|^{2}).
\end{eqnarray}

Similarly, the eigenvalues of  ${\cal W}_c$ are given as
\begin{eqnarray}
\lambda_1^{(c)} &=&\sum_{j=1,3}(\lambda
_{1}^{(a)}|x_{j}|^{2}+\lambda _{2}^{(a)}|y_{j}|^{2}), \\
\lambda_2^{(c)} &=&\sum_{j=2,4}(\lambda
_{1}^{(a)}|x_{j}|^{2}+\lambda _{2}^{(a)}|y_{j}|^{2}).
\end{eqnarray}

Then one has the sum of two separability coherences,
\beqa
{\cal S}_b+{\cal S}_c&=&2-2\lambda
_{1}^{(a)}(|x_{2}|^{2}+|x_{3}|^{2}+2|x_{4}|^{2}  )  \notag \\
&&-2\lambda _{2}^{(a)}(|y_{2}|^{2}+|y_{3}|^{2}+2|y_{4}|^{2}) \notag \\
&\le&2-2\lambda_{1}^{(a)} \sum_{j=2}^4|x_{j}|^{2} -2\lambda _{2}^{(a)}  \sum_{j=2}^4|y_{j}|^{2} \notag \\
&\le&2-2\lambda _{2}^{(a)} \sum_{j=2}^4(|x_{j}|^{2}+|y_{j}|^{2}). \label{Sab}
\eeqa

From the orthonormal conditions (\ref{condition1}) and (\ref{condition2}) it is easy to see that 
\begin{equation} 
1\geq |x_{1}|^{2}+|y_{1}|^{2},  \label{simplified-inequality}
\end{equation}
and when combining the two results (\ref{Sab}) and (\ref{simplified-inequality}), it is straightforward to obtain
\beqa
{\cal S}_b+{\cal S}_c&\le&2-2\lambda _{2}^{(a)} = 1 + {\cal S}_a,
\eeqa
which proves relation (\ref{inequality1}). The remaining inequalities (\ref{3inequality}) can be proved in exactly the same way. \\

\noindent\textbf{Acknowledgement:} \quad We acknowledge financial support from DARPA D19AP00042, and NSF grants PHY-1203931, PHY-1505189, and INSPIRE PHY-1539859.\\

\end{document}